\documentclass[12pt,thmsa]{article}

\usepackage{amsfonts}
\usepackage{graphicx}
\usepackage{epsfig}
\usepackage{graphics}

\makeatletter
\newcommand{\row}[1]%
{\mathord{\buildrel{\lower3pt%
\hbox{$\scriptscriptstyle\rightarrow$}}\over #1}}

\newcommand{\dyadic}[1]{\mathord{\dyadic@rrow{#1}}}
\newcommand{\dyadic@rrow}[1]{
\begin{picture}(12,12)(-1,0)
\put(-3,12){\makebox(0,0)[t]{$\scriptscriptstyle\downarrow$}}
\put(-3,13){\makebox(0,0)[l]{$\scriptscriptstyle\longrightarrow$}}
\put(5,0){\makebox(0,0)[b]{$#1$}}
\end{picture}
}

\newcommand{\bra}[1]{\bigl\langle #1 \bigr|}
\newcommand{\ket}[1]{\bigl| #1 \bigr\rangle}

\begin{document}
\begin{center}
{\large Dynamics of encrypted information in the presence of
imperfect operations}

N. Metwally \\[0pt]

Math. Dept., College of Science, University of Bahrain, Bahrain. \\[0pt]
E.mail: Nmetwally$@$gmail.com
\end{center}
 \begin{abstract}
The original dense coding protocol is achieved via quantum channel
generated  between a single Cooper pair and a cavity. The dynamics
of the coded and decoded information are investigated for
different values of the channel's parameters. The efficiency of
this channel  for  coding  and decoding  information depends on
the initial state settings of the Cooper pair. It is shown that,
these information increase as the detuning parameter increases or
the number of photons inside the cavity decreases. The coded and
decoded information increase as the ratio of the capacities
between the box and gate decreases. In the presence of imperfect
operation, the sensitivity of the information  to the phase error
is much larger than the bit flip error.

\bigskip
{\bf Keywords:} Dense coding,  Channels, Local and non-local information.\\

\end{abstract}

\section{Introduction}

One of the most obstacles of the quantum information processing is
sending  quantum and classical information safely. So, studying
the behavior of information in a noise circumstance is one of most
important tasks in the context of quantum communication
\cite{Nil}. Quantum dense coding, which was first proposed  by
Bennett and Wiesner\cite{Ben}, enables the communication of two
bits of classical information with the transmission of  qubit.
Therefore it has an important  applications in secure quantum
communication. To achieve a perfect quantum dense coding one needs
maximum entangled channels and perfect local operations  which are
difficult to be performed on the real world \cite{qing}.

 Therefore there are some
efforts have been done to investigate the possibility of
performing the dense coding  protocol in imperfect circumstances.
For example, overcoming a limitation of deterministic dense coding
with a non-maximally entangled initial state has been investigated
by Bourdon and  Gerjuoy \cite{Bour}. The dynamics and the
robustness of the coded information over the noise Bloch channels
are investigated in \cite{Met}.
 Xi-Han Li and et. al.\cite{Han} have used the
idea of quantum dense coding to investigate the robust quantum key
distribution protocols against two kinds of collective noise.

In this contribution, we  discuss the dynamics of the coded and
decoded information when the users apply imperfect operations
during the coding process.  Due to their potential in quantum
information processing  we use the generated entangled state
between a Cooper pair and a cavity mode as a quantum channel
\cite{Chi,Wall}. Metwally and et.al.\cite{Met1} have investigated
the  entangled properties of this state and discussed the
possibility of using it as quantum channel for quantum
teleportation .

 The paper is organized as follows: In Sec.$2$, an analytical solution of
 the suggested model is introduced. The coding protocol is
 performed, where the local operations  are archived perfectly in Sec.3. In Sec.4, the dynamics of
the coded and decoded information
 in the presence of two different types of noise are investigated.
 Due to this noise operations we quantify the disturbance of the
 decoded information in Sec.5. Finally, Sec.$6$ is devoted to discuss the
 results.

\section{The Model}
This model consists of  suberconducting box with a
low-capacitance, $C_{J}$ and Josephson junction, $E_{J}$ biased by
a classical voltage, $V_{g}$ through a gate capacitance, $C_{g}$
placed inside a single mode microwave cavity \cite{Nak,Tsai,Mig}.
If  the gate capacitance is screened from the quantized radiation
field, then junction-field Hamiltonian, in the interaction picture
can be written as,

\begin{equation}\label{H1}
\mathcal{H}_{c}=4E_{c}(n-n_{g})^{2}-E_{j}\cos\phi ,  \label{sys1}
\end{equation}%
where, $E_{c}=\frac{1}{2}e^{2}\left( C_{J}+C_{g}\right)$ is the
charging energy, $E_{J}=\frac{1}{2}\frac{\hbar}{e} I_c$  is the
Joesphson coupling energy, $e$ is the charge of the electron,
$n_{g}=\frac{1}{2}\frac{V_{g}}{e}C_g$ is the dimensionless gate
charge, $n$ is the number operators of excess Cooper pair on the
island and $\phi$ is the phase operator \cite{Mig,Rod}. If the
Josephon coupling energy, $E_{j}$
 is much smaller than the charging energy i.e
$E_{j}<<E_{c}$,  and the temperature is low enough, then  the
Humilation (\ref{H1}) becomes

\begin{equation}
\mathcal{H}_{c}=-\frac{1}{2}B_{z}\sigma
_{z}-\frac{1}{2}B_{x}\sigma _{x},
\end{equation}%
where $B_{z}=-\left( 2n-1\right) E_{cl}$, $E_{cl}$ is the electric
energy, $B_{x}=E_{j}$  and $\sigma _{x},\sigma _{y},\sigma _{z}$
are Pauli matrices. This Cooper pair can be viewed as an atom with
large dipole moment coupled to microwave frequency photons in a
quasi-one-dimensional transmission line cavity (a coplanar
waveguide resonator). The combined Hamiltonian for the Cooper
qubit and transmission line cavity is given by \cite{Sch,You},

\begin{equation}
\mathcal{H}=\varpi a^{\dagger }a+\varpi _{c}\sigma _{z}-\lambda (\mu -\nu \sigma _{z})+%
\sqrt{1-\nu ^{2}}\sigma _{x}(a^{\dagger }+a),  \label{Sys2}
\end{equation}%
where,  $\omega$  is the cavity resonance frequency,
$\omega_c=\sqrt{E_{j}^{2}+16E_{c}^{2}\left( 2n_{g}-1\right) ^{2}}$
 is the transition frequency of the Cooper  qubit, $\lambda
=\frac{\sqrt{C_{j}}}{C_{g+C_{J}}}\sqrt{\frac{1}{2}\frac{\varpi
}{\hbar }e^{2}}$ is coupling strength of resonator to the Cooper
qubit, $\mu =1-n_{g}$, $\nu =\cos\theta$ and $\theta =-\arctan
\Bigl(\frac{1}{E_{c}}\frac{E_{j}}{2n_{g}-1}\Bigr)$ is mixing
angle.

Assume that  the system is initially prepared in the state
$\ket{\psi _{s}(0)} =\ket{\psi_c}\otimes\ket{n}$, where
$\ket{\psi_c}=\frac{1}{\sqrt{2}}(\ket{e}+\ket{g})$ is the initial
state of the Cooper qubit,  while   $\ket{n}$ represents the
initial state of the  field. The time evolution of the initial
system is given by,
\begin{eqnarray}\label{final}
\ket{\psi _{s}(t)}& =&e^{-i\mathcal{H}t}\ket{\psi _{s}(0)} ,
\nonumber\\
&=&c_1\ket{n,e}+c_2\ket{n-1,e}+c_3\ket{n,g}+c_4\ket{n+1,g},
\end{eqnarray}
where,
\begin{eqnarray}\label{cof}
c_1&=&\cos\gamma_{n+1}
\tau-\frac{i\delta}{\gamma_{n+1}}\sin\gamma_{n+1} \tau,\quad c_2=
i\frac{\delta}{\gamma_{n+1}}\sqrt{n-1}\sin\gamma_{n+1}\tau,
\nonumber\\
c_3&=&\cos\mu_n \tau+i\frac{\delta}{\mu_n}\sin\mu_n \tau,
\quad\hspace{0.8cm} c_4=
-i\frac{\delta}{\gamma_{n+1}}\sqrt{n+1}\sin\gamma_{n+1}\tau,
\end{eqnarray}
and  $\delta=\frac{\Delta}{2\lambda}$,
 $\Delta =E_{j}-\varpi $ is the detuning parameter between the
Josephson energy and the cavity field frequency,
$\gamma_n=\sqrt{\delta^2+n}$ and $\tau=\lambda t$ is the scaled
time.

The entangled and separable behavior of this system are
investigated in \cite{Met1}, where different initial states
settings  are consider. The  effect of the Cooper pair and the
cavity's parameters on these phenomena is discussed. Also, the
possibility of using the generated entangled state between the
field and Cooper pair to achieve  quantum teleportation is studied
\cite{Met1}. In this context, we sheet the light on the dynamics
of the coded and decoded information via this type of entangled
states. Moreover, we investigate these  phenomena when the used
local operations are imperfect.

\section{Quantum coding via perfect operation}
For the sake of simplicity, we consider that the initial state of
the system is $\ket{\psi_s(0)}=\ket{n,e}$. For this initial state
setting,  the final state of the system is given by,
\begin{equation}\label{StCod}
\ket{\psi_s(t)}=c_1\ket{n,e}+c_4\ket{n+1,g},
\end{equation}
where $c_1$ and $c_4$ are given from (\ref{cof}). This state
represents the entangled channel between the users Alice and Bob.
 To perform quantum coding protocol, the users need
to apply a group of local operations. Due to the noise, these
operations can not be performed perfectly. Therefore, our aim in
this section is investigating the dynamics of the coded and
decoded information when the operations are archived correctly. To
show this idea, we implement the original dense coding protocol
which has been proposed by Bennett and Wienser \cite{Ben}.  This
protocol is described as follows:

\begin{enumerate}
\item Alice encodes two classical bits by using one of local
unitary operations. If Alice applies these unitary operations
randomly with probability $p_i$, then she codes the information in
the state,
\begin{equation}\label{cod1}
\rho_{c}=\sum_{i=0}^{3}\Bigl\{p_i\sigma_{1i}\otimes
I_2\rho_s(t)\sigma_{1i}\otimes I_2\Bigr\},
\end{equation}
where,
\begin{equation}\label{system}
\rho_s(t)=c_1^2\ket{e,n}\bra{e,n}+c_1c_4^*\ket{e,n}\bra{g,n+1}+c_4c_1^*\ket{g,n+1}\bra{n,e}+c_4^2\ket{g,n+1}\bra{g,n+1},
\end{equation}
 represents  the quantum channel between the users Alice and Bob and $\sigma_{1i}= I_1,
\sigma_{1x},\sigma_{1y}$ and $\sigma_{1z}$,
 are the unitary operators, which represent the local operations for
 Alice's  qubit and $I_2$ is the identity operator for  Bob's
qubit. The coded information is given by ,
\begin{equation}
I_{cod}=-c_1^2log_2c_1^2-c_4^2log_2c_4^2,
\end{equation}
where it is assumed that Alice performs the local operations with
an equal probability i.e., $p_i=\frac{1}{4}, j=0...3$.
 \item Alice sends her qubit to Bob, who makes  joint
measurements on the two qubits. The maximum amount of information
which Bob can extract from Alice's message is bounded by,
\begin{eqnarray}
I_{Bob}&=&\mathcal{S}\Bigl(\sum_{j=0}^{j=3}p_j\rho(t)\Bigr)-\sum_{i=0}^{j=3}p_i\mathcal{S}(\rho(t)),
\end{eqnarray}
where $\mathcal{S}(.) $ is the von Neumann Entropy.
\end{enumerate}
\begin{figure}[t!]
  \begin{center}
 \includegraphics[width=19pc,height=14pc]{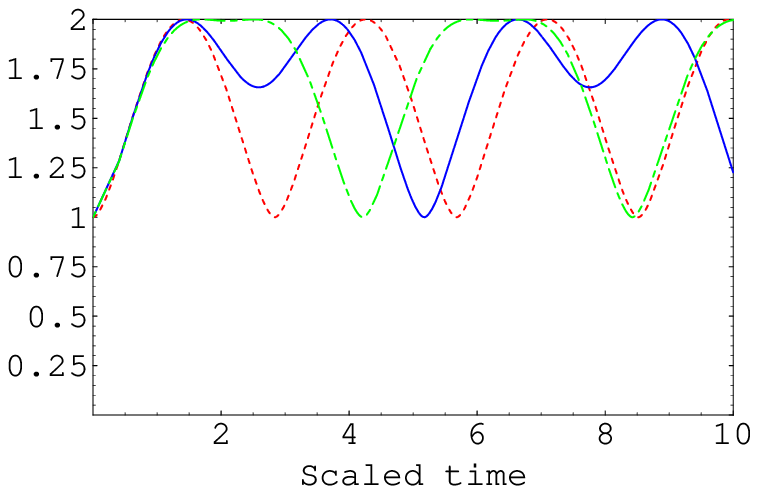}
 \includegraphics[width=19pc,height=14pc]{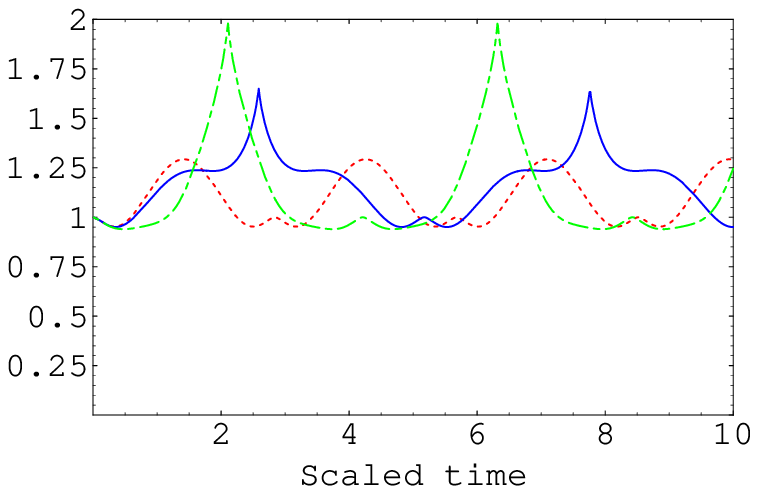}
\put(-460,90){$I_{cod}$}
 \put(-222,90){$I_{Bob}$}
 \put(-263,32){$(a)$}
 \put(-28,32){$(b)$}
    \caption{The dynamics of the information in the presence of perfect local operations where
     $n=2$, the ratio  $\kappa=\frac{\sqrt{C_{j}}}{C_{g+C_{J}}}=\frac{5}{2}$ and $\Delta=0,0.5,1$ for the dot, solid and dash-dot curves respectively
     (a)coded information, $I_{cod}$ (b) decoded information, $I_{Bob}$.}
  \end{center}
\end{figure}
\begin{figure}
  \begin{center}
 \includegraphics[width=19pc,height=14pc]{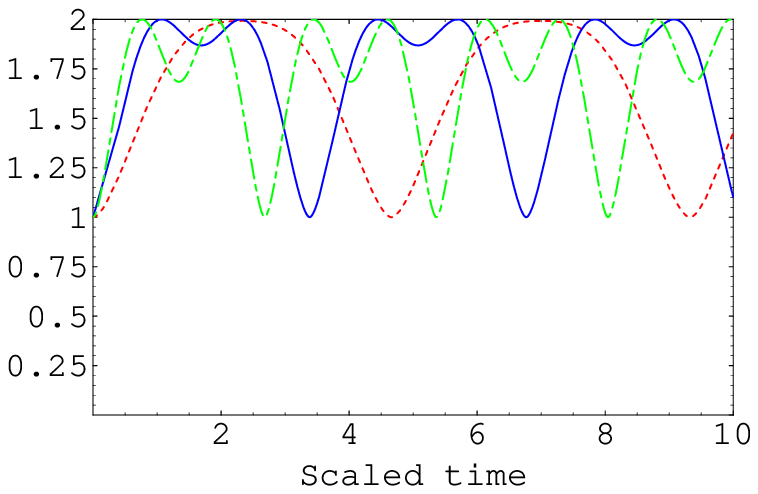}
 \includegraphics[width=19pc,height=14pc]{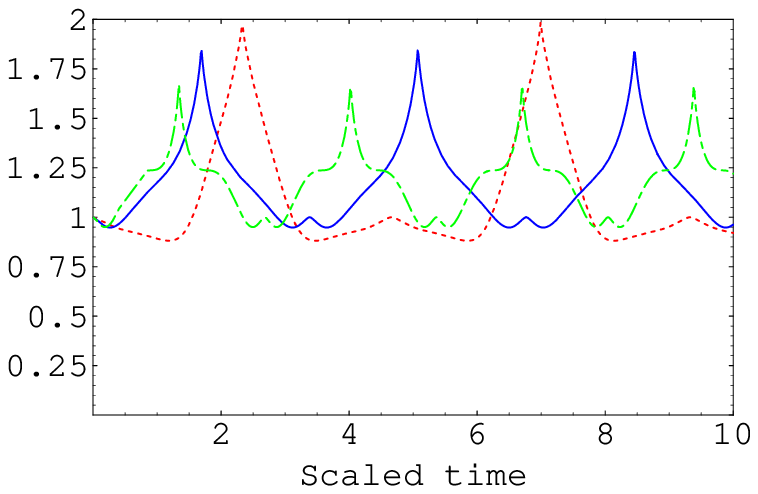}
\put(-460,90){$I_{cod}$}
 \put(-222,90){$I_{Bob}$}
\put(-263,32){$(a)$}
 \put(-28,32){$(b)$}
    \caption{The same as Fig.(1), but $n=1,5$ and $10$ for the dot, solid, dash-dot curves respectively
    and
    $\Delta=1$.}
  \end{center}
\end{figure}

The dynamics of the coded information, $I_{cod}$ and the decoded
information, $I_{Bob}$ are investigated in Fig.(1) for different
values of the detuning parameter, where we fixed the values of
other parameters. For small values of $\Delta$, the coded
information increases gradually to reach its maximum values and
then decreases smoothly. As one increases $\Delta$, the number of
oscillations decreases and $I_{cod}$ remain  maximum for a longer
time. This is clear by comparing the dot curve where we set
$\Delta=0$ and the dash-dot curve ($\Delta=1)$ in Fig.(1a). The
effect of the detuning parameter on the decoded information is
displayed in Fig.(1b). In general the amount of information which
Bob can extract from the sending information increases as one
increases the detuning parameter. However, the maximum extracted
information depends on the value of $\Delta$.

In Fig.(2), we investigate the effect of  the   number of photons,
$n$ on the dynamics of the coded and decoded information. Fig.(2a)
shows the behavior of $I_{cod}$ for different values of $n$  while
the detuning parameter, $\Delta=1$. The general behavior of
$I_{cod}$ is the same as that depicted in Fig.(1a). However the
number of oscillations increases as one increases $n$. The
dynamics of the decoded information is displayed in Fig.(2b),
where the maximum extracted information increases as one decreases
the number of photons inside the cavity.

\begin{figure}[t!]
  \begin{center}
 \includegraphics[width=19pc,height=13pc]{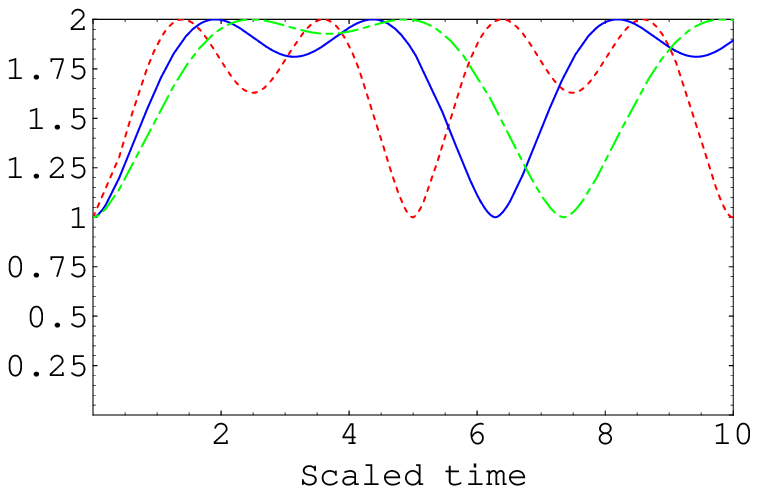}
 \includegraphics[width=19pc,height=13pc]{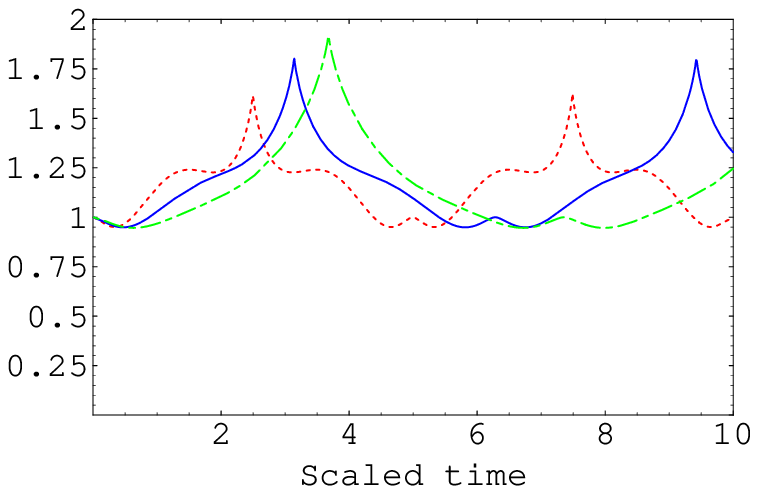}
\put(-460,90){$I_{cod}$}
 \put(-222,90){$I_{Bob}$}
 \put(-263,32){$(a)$}
 \put(-28,32){$(b)$}
    \caption{The same as Fig.(1), but the ratio, $\kappa=\frac{1}{2},\frac{1}{3}$ and $\frac{1}{4}$ for the dot, solid and dash
    dot curves respectively , with $\Delta=0.5$ and $n=2$.}
  \end{center}
\end{figure}
In this context,it is important to investigate the effect of the
parameter $\kappa$, which represents the ratio between the
capacities of the Cooper pair box and the gate.  The dynamics of
the coded and decoded information for different values of $\kappa$
are shown in Fig.(3). It is clear that as $\kappa$ decreases, the
lower bound of coded information shifted to the right. This means
that the coded information  remains maximum for larger interval of
time. This behavior is clearly seen by comparing Fig.(1a) and
Fig.(3a). However as shown in Fig.(3b), for smaller values of
$\kappa$, the maximum values of  the decoded information increases
and shifted to the right.

Let us end this section by considering another initial state
settings of the system. Assume that the Cooper qubit is prepared
initially in the ground state $\ket{g}$. Then the initial state of
the total system is $\rho_s(0)=\ket{n,g}\bra{n,g}$. The dynamics
evolution of this system is given by
\begin{equation}\label{gr}
\rho_s(t)=|c_2|^2\ket{n-1,e}\bra{n-1,e}+c_2c^*_3\ket{n-1}\bra{n,g}+c_3c_2^*\ket{n,g}\bra{n-1,e}+|c_3|^2\ket{n,g}\bra{n,g},
\end{equation}
where $c_2$  and $c_3$ are given  from (\ref{cof}). The users
Alice and Bob use the state(\ref{gr}) as a quantum channel to
perform the original dense coding protocol \cite{Ben}.

\begin{figure}
  \begin{center}
 \includegraphics[width=19pc,height=13pc]{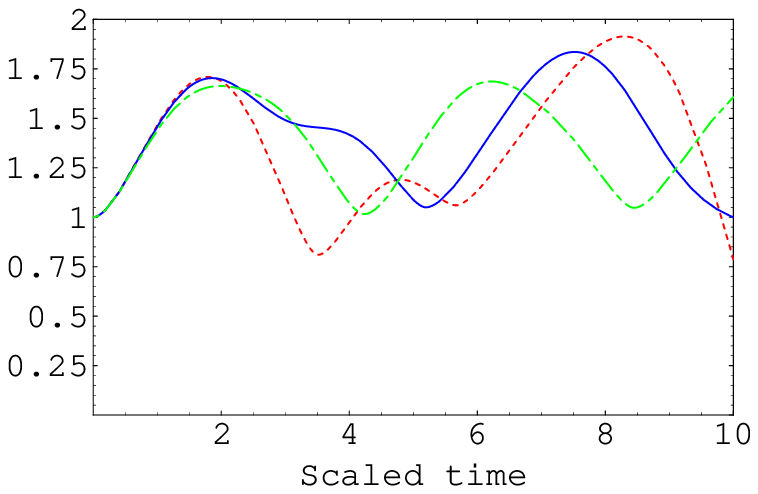}
 \includegraphics[width=19pc,height=13pc]{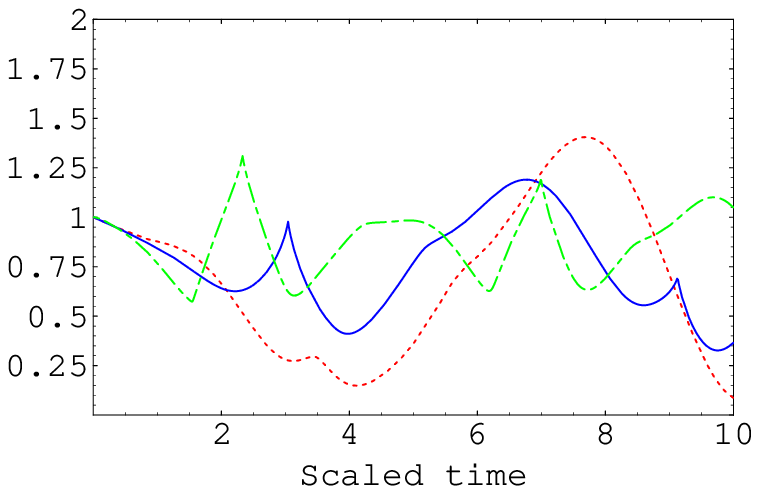}
\put(-460,90){$I_{cod}$}
 \put(-222,90){$I_{Bob}$}
 \put(-263,32){$(a)$}
 \put(-28,32){$(b)$}
    \caption{The same as Fig.(1), but the system is initially prepared in the state $\psi_s(0)=\ket{n,g}$.}
  \end{center}
\end{figure}

Fig.(4) describes the dynamics of the coded and decoded
information for different values of the detuning parameter, which
is enough to show  the effect of different initial state settings
on the dynamics of information. The effect of the detuning
parameter on the coded and decoded information is the same as that
shown in Fig.(1). From Fig.(4) and Fig.(1), it is clear that
coding information in a system prepared initially in an excited
state, $\ket{e}$ is much better than using the ground state
$\ket{g}$.

From our finding, the users can coded their information with high
rate by increasing the detuning parameter or  decreasing the
number of photons inside the cavity. However the number of
oscillations increases for smaller values of the detuning
parameter and larger values of the number of photons. Coding
information in a system of Cooper pair prepared initially in an
exited state  is much better than prepared it in ground state.

\section{Quantum coding via imperfect operations}

In this section, the dynamics of the coded and decoded information
are investigated when some of the local operations are performed
imperfectly during the coding process. In this context, we assume
that there are two types of imperfect operations: the {\it first},
Alice performs the bit flip operation with a certain probability,
while the {\it second}, instead of performing  bit flip, Alice
performs a phase flip operation.

\subsection{First type of imperfect operation}
Let us assume that during the coding process, Alice applies the
operation $\sigma_x$ successful with probability $q$  and fails
with probability $(1-q)$. In this case, the information is coded
in the state,

\begin{equation}\label{cod1}
\rho_{c}=\frac{q}{4}\sigma_{1x}\rho_s(t)\sigma_{1x}+\frac{1-q}{4}\rho_s(t)+\frac{1}{4}\sum_{i=0}^{2}\Bigl\{\sigma_{1i}\otimes
I_2\rho_s(t)\sigma_{1i}\otimes I_2\Bigr\},
\end{equation}
where $\sigma_{1i}=I_1,\sigma_{1y}$ and $\sigma_{1z}$ are the
other local operations which are performed perfectly.

Fig.(5), shows the behavior of the coded and decoded information
where  Alice  performed the bit flip operation correctly with
probability $q=0.1$. From Fig(5a), it is clear that the coded
information, $I_{cod}$ decreases very fast for small values of the
detuning parameter. However as one increases $\Delta$, the coded
information decreases gradually and the minimum value is larger
than that depicted for small values of $\Delta$. The behavior of
the decoded information, $I_{Bob}$ is displayed in Fig.(5b), where
the amount of the  extracted information increases as one increase
$\Delta$. Also, the minimum values of $I_{Bob}$ increases as the
detuning parameter increases.

\begin{figure}[b!]
  \begin{center}
  \includegraphics[width=19pc,height=14pc]{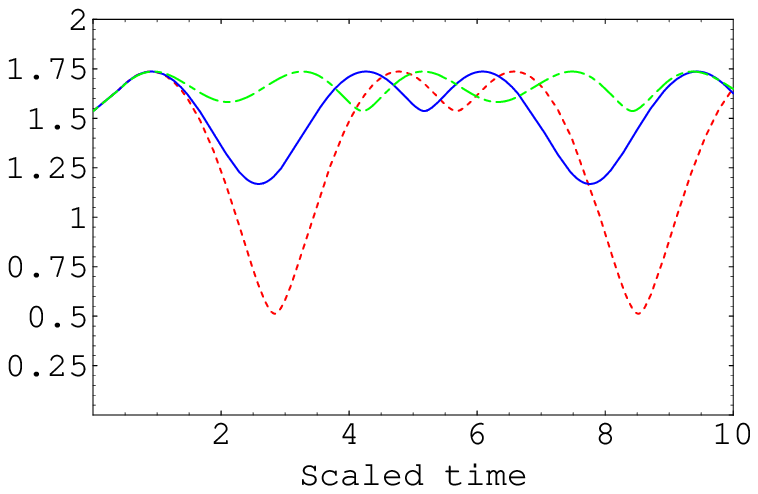}
  \includegraphics[width=19pc,height=14pc]{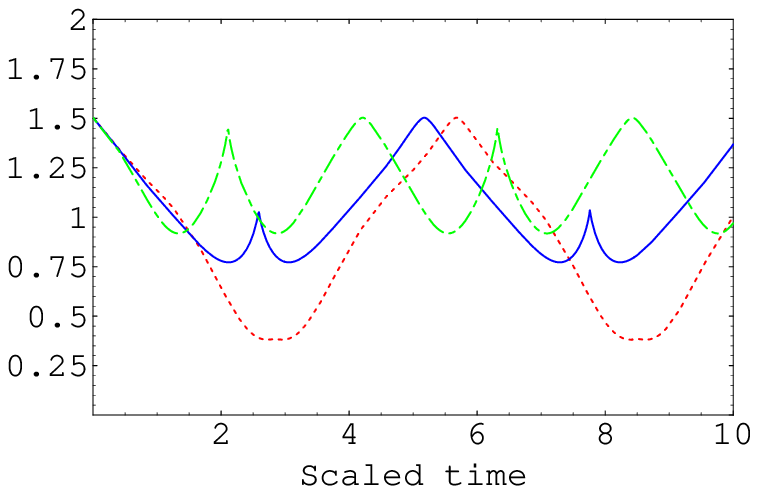}
\put(-460,90){$I_{cod}$}
 \put(-222,90){$I_{Bob}$}
 \put(-263,32){$(a)$}
 \put(-28,32){$(b)$}
    \caption{The  same as Fig.(1) but Alice performs the bit flip operation
    with a probability $q=0.1$ and fils with probability $q=0.9$ (first type of imperfect operation).}
  \end{center}
\end{figure}

\begin{figure}
  \begin{center}
  \includegraphics[width=19pc,height=14pc]{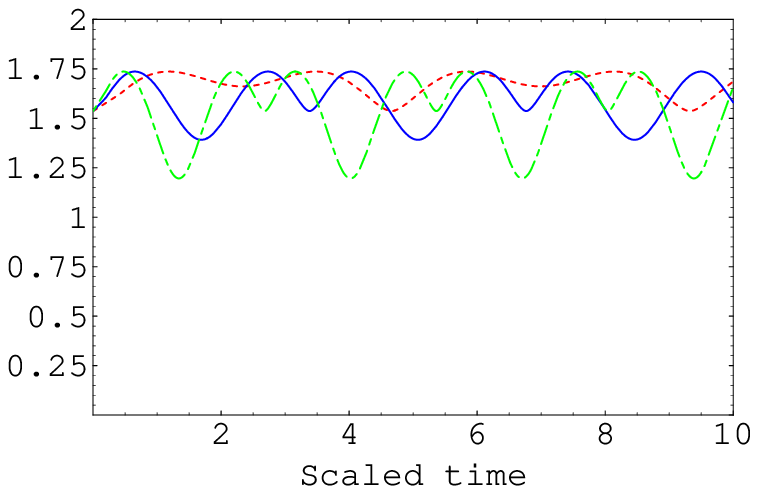}
\includegraphics[width=19pc,height=14pc]{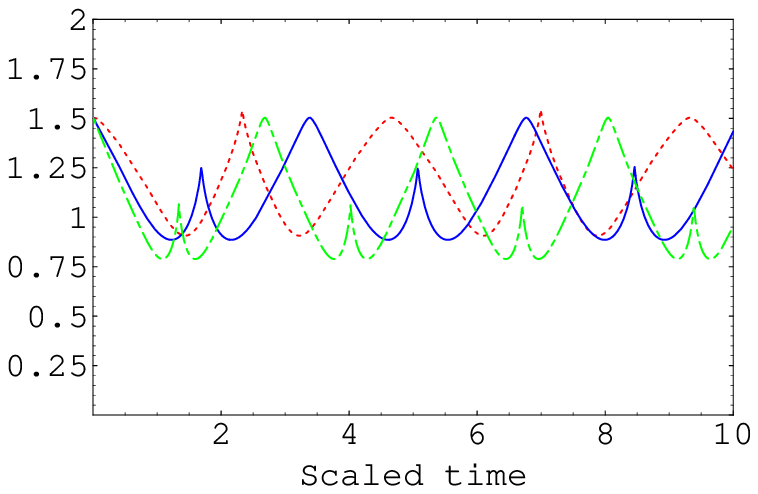}
\put(-460,90){$I_{cod}$}
 \put(-222,90){$I_{Bob}$}
 \put(-263,32){$(a)$}
 \put(-28,32){$(b)$}
    \caption{The  same as Fig.(5) but for different values of the number of photons inside the
    cavity. The dot, solid and dash-dot curves
      $n=1,5,10$ respectively. }
  \end{center}
\end{figure}

The effect of the number of photons, $n$ inside the cavity is
shown in Fig.(6).  It is clear that, as one increases $n$, the
coded information decreases. The minimum value is much larger than
that depicted in Fig.(5a). This shows that, the amount of
information which is coded in the system (\ref{system})  can be
improved by increasing the detuning parameter or decreasing the
number of photons inside the cavity. The behavior of the decoded
information is displayed in Fig.(5b). In general the behavior of
$I_{Bob}$ is similar to that shown in Fig.(5b). However the number
of fluctuations increases as one increases $n$ and the minimum
value of the decoded information increases for larger values of
$n$.

\subsection{Second type of imperfect operation}
We assume that  Alice applies the local operation $\sigma_z$(phase
flip), instead of  $\sigma_x$ (bit flip) during the coding
process. In this case, the information is coded in the state,
\begin{equation}\label{nz}
\rho_c=\frac{1}{2}\sigma_{1z}\otimes
I_2\rho_s(t)\sigma_{1z}\otimes
I_2+\frac{1}{4}\sum_{i=0}^{1}\Bigl\{\sigma_{1i}\otimes
I_2\rho_s(t)\sigma_{1i}\otimes I_2\Bigr\},
\end{equation}
where $\sigma_{1i}=I_1,\sigma_{1y}$. The amount of information
which is coded in the state (\ref{nz}) is given by

\begin{equation}
I_{cod}=-\frac{3}{4}\Big[(c_1^2log_2\frac{3c_1^2}{4}+c_4^2log_2\frac{3c_4^2}{4}+\frac{c_1^2+c_4^2}{3}
log_2\frac{c_1^2+c_4^2}{4}\Bigr]
\end{equation}
\begin{figure}[t!]
  \begin{center}
\includegraphics[width=19pc,height=14pc]{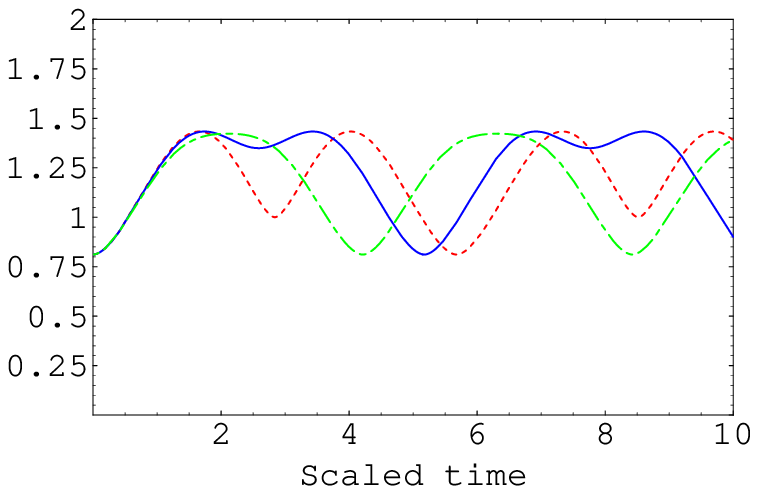}
 \includegraphics[width=19pc,height=14pc]{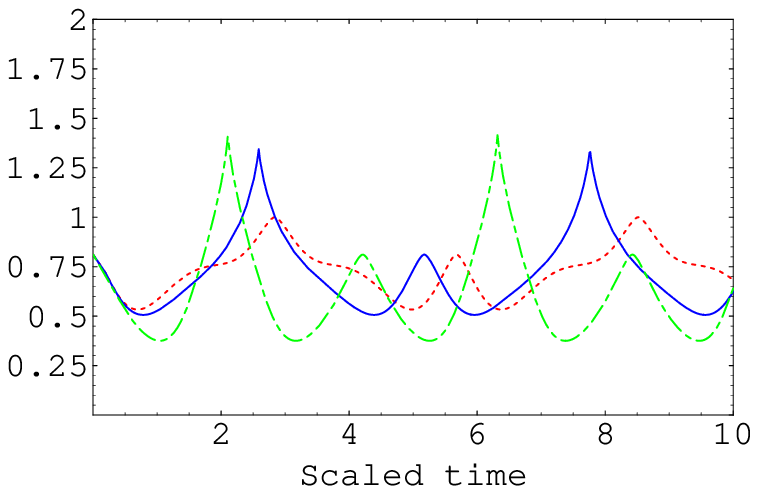}
\put(-460,90){$I_{cod}$}
 \put(-222,90){$I_{Bob}$}
 \put(-263,32){$(a)$}
 \put(-28,32){$(b)$}
    \caption{The  same as Fig.(1) but the Alice applies phase flip operation instead of the bit flip operation
 (second  type of imperfect operation).}
  \end{center}
\end{figure}
\begin{figure}
  \begin{center}
 \includegraphics[width=19pc,height=14pc]{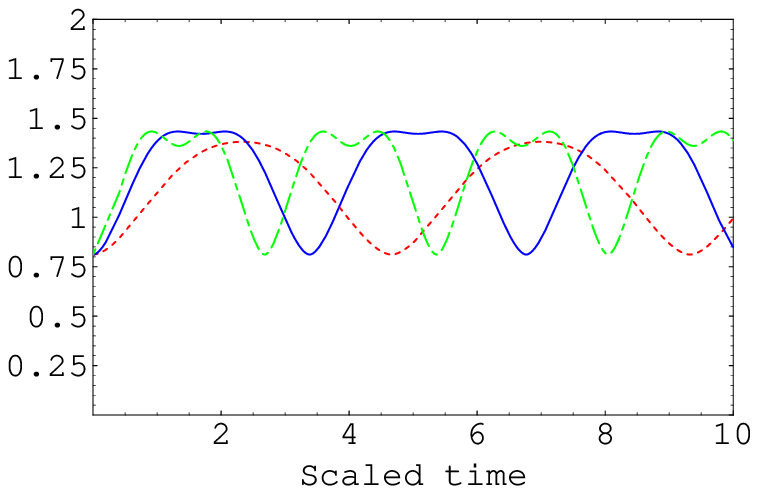}
 \includegraphics[width=19pc,height=14pc]{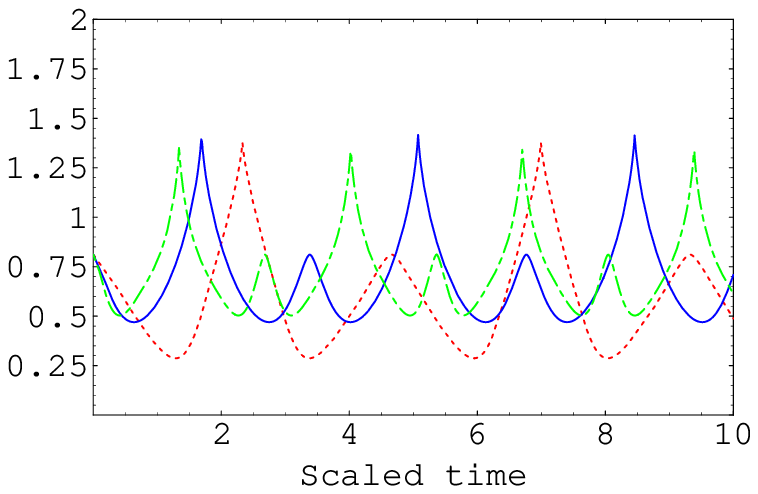}
\put(-460,90){$I_{cod}$}
 \put(-222,90){$I_{Bob}$}
 \put(-263,32){$(a)$}
 \put(-28,32){$(b)$}
    \caption{The  same as Fig.(2) but Alice applies phase flip operation instead of the bit flip operation
 (second  type of imperfect operation).}
  \end{center}
\end{figure}
Fig.(7) shows the dynamics of the coded  information in the
presence of imperfect operation for different values of the
detuning parameter $\Delta$, while the other parameters are fixed.
From Fig.(7a), it is clear that the  general behavior is the
similar to that shown in Fig.(5a), but the upper bound is smaller.
 Fig.(7b) displays the effect of $\Delta$,
on the dynamics of the decoded information. In this case, the
upper and lower bounds are smaller than that shown in Fig.(6b).
 The effect of the number of photons,
$n$ on the dynamics of the coded and decoded information is
described in Fig.(8). In general these phenomena behaves as that
shown in Fig.(7), only the number of oscillations increases as $n$
increases.
\begin{figure}[t]
  \begin{center}
 \includegraphics[width=19pc,height=14pc]{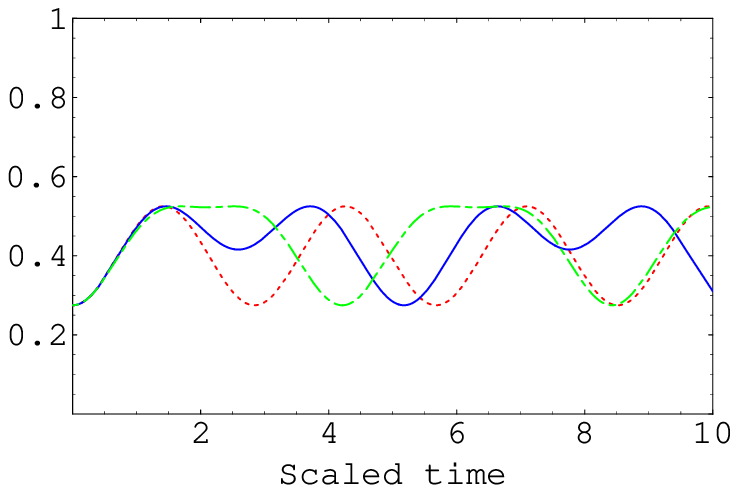}
 \includegraphics[width=19pc,height=14pc]{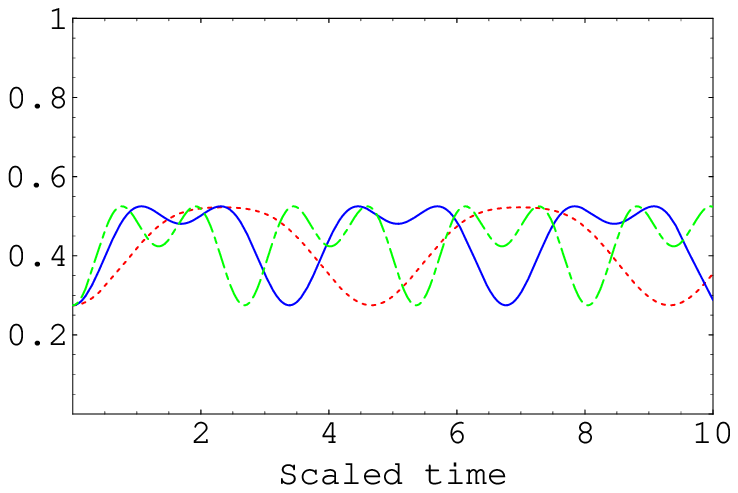}
 \put(-265,32){$(a)$}
 \put(-35,32){$(b)$}\\
 \includegraphics[width=19pc,height=14pc]{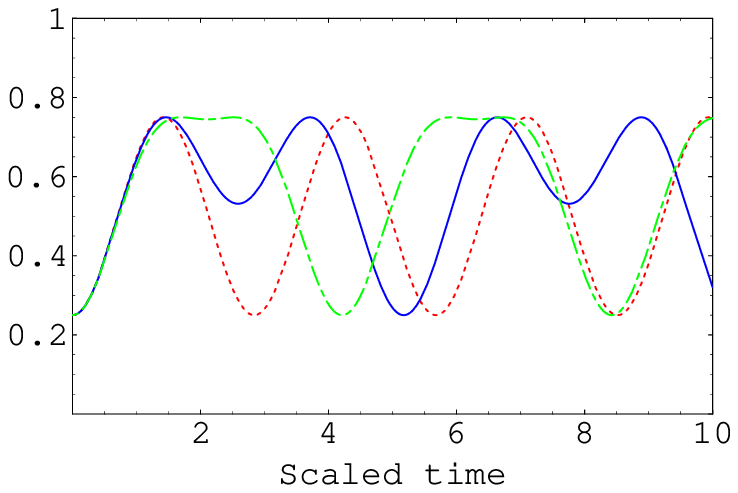}
\includegraphics[width=19pc,height=14pc]{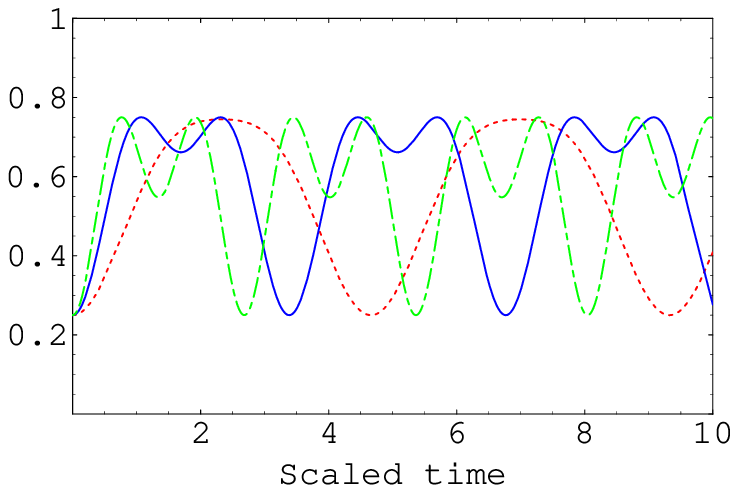}
\put(-462,90){$\mathcal{D}$}
 \put(-224,90){$\mathcal{D}$}
 \put(-265,32){$(c)$}
 \put(-35,32){$(d)$}
    \caption{Figs.($a\&b$)  represent the disturbance
    when the user applies the first type of imperfect operations,
    while Figs.($c\&d$)   for the
    second type of imperfect operations. In  Figs.(a$\&c$), the dot, solid and
    dash-dot for $\Delta=0,0.5$ and $1$ respectively, while in
    Figs.(b$\&d$)  $n=1,5$ and $10$ respectively.}
  \end{center}
\end{figure}

 Figs.(5$\&6$) and  Figs.(7$\&$8), the dynamics of information is
 very sensitive to the noisy operations during the coding process.
  This shows that the error in achieving the phase
bit operation has a larger effect than the error of applying the
bit flip operation on the dynamics of information.

\section{Disturbance}

In this section,  we  quantify how much the coded information is
disturbed or equivalently how much the initial coded  and final
decoded states are not coincide. This phenomena is called a
disturbance of information which is  defined as
\cite{Mac,Metwally1},
\begin{equation}
\mathcal{D}=1-\mathcal{F},
\end{equation}
where $\mathcal{F}$ is the average fidelity. The dynamics of the
disturbance is displayed in Fig.(9) for  different types of
imperfect local operations.
 It is clear that, when Alice applies
the flip operator imperfectly, first type of imperfect operation,
(Fig.9a$\&9b$) the upper bound of the disturbance  is much smaller
than that described for the second type of imperfect operation
(Fig9c$\&9d)$. This means that the average fidelity of the decoded
information in the presence of the first type of imperfect
operation  is much better than that for the second type. In other
words, the coded information is very sensitive to the phase error.

\section{Conclusion}
The dynamics of the encrypted information in a single Cooper pair
interacts with a cavity is investigated. The effect of the filed
and the Cooper pair's parameters on the dynamics of coded and
decoded information is discussed. We can see that as soon as the
interaction goes on the coded information increases, which means
that the generated entangled state  between the cavity mode and
the Cooper qubit can be used to code information with high
efficiency. This efficiency depends on the initial state settings
of the Cooper pair, where it is shown that the quantity of coded
and decoded information is larger if the Cooper  qubit  is
prepared initially in an excited state. Also, it is shown that the
users can improve the amount of the coded and decoded information
by increasing the detuning parameter or decreasing the number of
photons. However the amount of the decoded information is much
smaller than the coded information. Therefore there are some
information has been lose during the transforming  process.

The dynamics of these phenomena are  investigated in the presence
of local imperfect operations, where it is assumed that during the
coding process one of the local operations is performed in
imperfectly. We consider that the Alice  applies the bit flip
operation with a probability $q$ and files with a probability
(1-q). In this case, the decoded information decreases very fast
for small values of the detuning parameter, while decreases
smoothly for larger values of photons. However as one increases
the detuning parameter the travelling information suppressing the
type of the local environmental noise. However decreasing the
number of photons inside the cavity can resist the lose of
information.

For the  other imperfect operation, where the Alice applies the
phase flip operation instead of the bit flip operation, the upper
bound of the coded and decoded information is much smaller than
that depicted for the first type noise. This means that the
suppressing of the  travelling coded information  to this  to type
of noise is fragile. However the lower bound of the decoded
information for this type of noise is much smaller than the first
type of noise. This shows that the error in achieving the phase
bit operation has a larger effect than the error of applying the
bit flip operation, namely the coded and decoded information are
very  sensitive to the phase operations.

Finally, the lost information due to the imperfect operation is
quantified.  It is shown that the efficiency of the channel
between the users decreases, where there is a large disturbance of
the information  during the coded and decoded processes. However
this disturbance due  to the error in the achieving the bit flip
operation is smaller than that shown for the phase flip error.

\end{document}